\documentclass[]{pasj02} 
\usepackage[switch,mathlines]{lineno} 
\usepackage{graphicx}
\usepackage{siunitx}

\jyear{2026}
\Received{}
\Accepted{}

\begin{document}

\title{One-week optical observations of pulsed emission from the Crab pulsar with IMONY on the 3.8 m Seimei telescope
\thanks{
This is a pre-copyedited, author-produced version of an article accepted for "publication in Publications of the Astronomical Society of Japan" following peer review.
The version of record (PASJ, 2026, Vol. xx, p. xxx) is available online at [https://doi.org/10.1093/pasj/psag026].}
}

\author{
    Kazuaki \textsc{Hashiyama},\altaffilmark{1}\altemailmark\orcid{0009-0001-7459-1670}\email{kazuaki.hashiyama@nao.ac.jp}
    Takeshi \textsc{Nakamori},\altaffilmark{2}\altemailmark\orcid{0000-0002-7308-2356}\email{nakamori@sci.kj.yamagata-u.ac.jp} 
    Anju \textsc{Sato},\altaffilmark{2}
    Mana \textsc{Hasebe},\altaffilmark{2}
    Miu \textsc{Maeshiro},\altaffilmark{2} \\
    Rin \textsc{Sato},\altaffilmark{2}
    Tomohiro \textsc{Sato},\altaffilmark{2}
    Masaru \textsc{Kino},\altaffilmark{3}
    Kazuhiro \textsc{Takefuji},\altaffilmark{4}
    Toshio \textsc{Terasawa},\altaffilmark{5} \\
    Koji S. \textsc{Kawabata},\altaffilmark{6,7}
    Tatsuya \textsc{Nakaoka},\altaffilmark{6}
    Dai \textsc{Takei},\altaffilmark{8,9}
    Masayoshi \textsc{Shoji},\altaffilmark{10}
    Shota \textsc{Kisaka},\altaffilmark{7} \\
    and Kazuki \textsc{Ueno}\altaffilmark{11}
}
\altaffiltext{1}{Mizusawa VLBI Observatory, National Astronomical Observatory of Japan, 2-12 Hoshigaoka, Mizusawa, Oshu, Iwate 023-0861, Japan}
\altaffiltext{2}{Faculty of Science, Yamagata University, 1-4-12 Kojirakawa, Yamagata 990-8560, Japan}
\altaffiltext{3}{Okayama Observatory, Kyoto University, 3037-5 Honjo, Kamogata-cho, Asakuchi, Okayama 710-0232, Japan}
\altaffiltext{4}{Usuda Deep Space Center, Japan Aerospace Exploration Agency, 1831-6 Omagari, Kamiodagiri, Saku, Nagano 384-0306, Japan}
\altaffiltext{5}{Institute of Cosmic-Ray Research, University of Tokyo, 5-1-5 Kashiwanoha, Kashiwa, Chiba 277-8582, Japan}
\altaffiltext{6}{Hiroshima Astrophysical Science Center, Hiroshima University, 1-3-1 Kagamiyama, Higashi-Hiroshima, Hiroshima 739-8526, Japan}
\altaffiltext{7}{Graduate School of Advanced Science and Engineering, Hiroshima University, 1-3-1 Kagamiyama, Higashi-Hiroshima, Hiroshima 739-8526, Japan}
\altaffiltext{8}{Daiphys Technologies LLC, 1-5-6 Kudan-Minami, Chiyoda, Tokyo 102-0074, Japan}
\altaffiltext{9}{Research Center for Measurement in Advanced Science, Rikkyo University, 3-34-1 Nishi-Ikebukuro, Toshima, Tokyo 171-8501, Japan}
\altaffiltext{10}{KEK, High Energy Accelerator Research Organization, Institute of Particle and Nuclear Studies, 1-1 Oho, Tsukuba, Ibaraki 305-0801, Japan}
\altaffiltext{11}{Graduate School of Science, Osaka University, 1-1 Machikaneyama, Toyonaka, Osaka 560-0043, Japan}




\maketitle

\begin{abstract}
We report our optical observations of the Crab pulsar using the Imager of MPPC-based Optical photoN counter from Yamagata (IMONY), a high-time-resolution photon-counting imager with 100~ns timing resolution, mounted on the 3.8~m Seimei telescope in Japan (f/D$\sim$6).
The detector format was upgraded from a $4\times4$ to an $8\times8$ GAPD array with larger pixels (100 to $200~\mathrm{\mu m}$), resulting in a $14\arcsec.5$ field of view on the Seimei telescope.
We conducted nightly optical observations for one week, including two nights of simultaneous optical and radio observations with the 64 m Usuda radio telescope.
Thanks to the large diameter of the Seimei telescope and the high time resolution of IMONY, we successfully detected optical Single Pulses (SPs) emitted in each rotation.
Moreover, we found an optical peak timing drift of $30\pm7.9~\mathrm{\mu s}$ over three days, with a significance of $3.9\sigma$.
The corresponding emission region size is $9.1~\mathrm{km}$, which is equivalent to 0.006 times the light cylinder radius of the Crab pulsar.
We ruled out the possibility of a pulsar glitch and suggested that the optical pulsed emission region of the Crab pulsar may fluctuate due to the spatial drift and variations in the magnetospheric caustics.



\end{abstract}


\section{Introduction} \label{sec:intro}

A pulsar is a fast rotating neutron star, and electromagnetic radiation from the pulsar is observed as a pulsed emission associated with its rotating period.
Although the emission mechanism of the pulsar is still under discussion, the radio emission is considered to be coherent radiation, while optical to gamma-ray emissions are thought to originate from incoherent radiation.
To improve our understanding of the pulsar emission mechanism, it is relevant to study how charged particles are produced and accelerated within the pulsar magnetosphere.
In recent years, Particle-In-Cell (PIC) simulations have provided detailed insights into these processes; however, due to computational limitations, they typically adopt simplified models of particle production and are restricted to limited regions of the magnetosphere (e.g., \citet{Philippov+22}).
Complementary magnetohydrodynamic (MHD) simulations (e.g., \citet{Tchekhovskoy+13}) can capture the global structure of the magnetosphere but lack the microphysical detail of PIC models.
By combining observational constraints on short-term variations of the pulsed emissions with insights from both PIC and MHD simulations, we can achieve a more comprehensive understanding of particle production and acceleration in pulsar magnetospheres.


PSR J0534+2200 (the Crab pulsar), which is located at the center of the Crab Nebula, is one of the brightest pulsars and a good candidate for observing the short-term variations of the pulsations.
The Crab pulsar emits the periodic electromagnetic pulses every approximately 34~ms, and they are detected in radio to gamma-ray energy bands \citep{Moffett+96,Rots+04,Aliu+08}.
These pulses are emitted twice in each rotation cycle, identified as a main pulse (MP) and an interpulse (IP).

During both the MP and IP phases, the Crab pulsar sometimes emits Giant Radio Pulses (GRPs), which are intense bursts of radio emission exceeding the typical intensity of normal radio pulses by more than one order of magnitude.
GRPs can exhibit extremely narrow pulse widths on the nanosecond scale, known as \textit{nanoshots} \citep{Hankins+07}, implying an emission region with a light travel size of $c\delta t\sim12~\mathrm{cm}$.
In addition, prior studies have found that optical (\citet{Shearer+03} and \citet{Strader+13}) and X-ray \citep{Enoto+21} pulsed emissions associated with GRPs are enhanced by approximately 3\% on average for the MP, with statistical significance exceeding $5\sigma$.
These correlations suggest that common physical processes are involved in GRP emission, at least from radio to X-ray bands.

In the optical band, Charge-Coupled Devices (CCDs) and Complementary Metal-Oxide-Semiconductor (CMOS) sensors are widely used for astronomical observations.
In recent years, Avalanche Photodiodes (APDs) have also been adopted as the Optical Pulsar TIMing Analyzer (OPTIMA) \citep{Kanbach+08} and the Asiago Quantum EYE (AquEYE) \citep{Zampieri+15} employ Geiger-mode Single Photon Avalanche Diodes (SPADs).
Also, a Microwave Kinetic Inductance Detector (MKID) made of superconducting material was adopted for the ARray Camera for Optical Near-infrared Spectroscopy (ARCONS) \citep{Mazin+13}.
These photometers have improved the time resolution from a few milliseconds to sub-nanosecond levels and successfully detected the optical pulse profile (phaseogram) of the Crab pulsar \citep{Shearer+03,Strader+13,Zampieri+14}.
Also, thanks to the high-time-resolution optical observations, the quasi-periodic mode change of the phaseogram was found \citep{Karpov+07}.
Moreover, \citet{Kanbach+08} successfully detected pulses per rotation, Single Pulses (SPs), using OPTIMA mounted on the 3.5 m Calar Alto telescope in 2002.
The observation was conducted as part of a performance evaluation of the instrument; therefore, statistical properties of optical SPs, which are important for studying the short-term variations of the optical pulsed emission, were not clarified.


We have developed the Imager of MPPC-based Optical photoN counter from Yamagata (IMONY), 
a high-time-resolution optical imager that timestamps individual photons with a 100~ns resolution. 
Its absolute timing accuracy is better than 200~ns, taking into account the jitter of the Global Navigation Satellite System (GNSS) and signal delays in the internal electronics \citep{Nakamori+21,Nakamori+25}.
The prototype of IMONY \citep{Nakamori+21} used a monolithic $4\times4$ GAPD array with a $100~\mathrm{\mu m}$ pitch, fabricated as a customized MPPC by Hamamatsu.
Unlike a conventional MPPC/SiPM, which provides a single summed output from thousands of microcells, our sensor comprises only a $4\times4$ microcell structure and provides independent readout for each microcell.
Hereafter, we refer to an individual microcell as a "pixel".
We mounted the prototype on the 1.5 m Kanata telescope (Hiroshima, Japan) and demonstrated its performance by detecting the Crab pulsar \citep{Nakamori+25}.
However, the field of view (FoV) is limited as comparable to the Point Spread Function (PSF) due to the limited number of pixels.
Subsequently, we have addressed this limit by expanding the pixel array to $8\times8$ with a $200~\mathrm{\mu m}$ pitch and upgrading the signal processing system.
We then installed IMONY on the 3.8 m Seimei telescope (Okayama, Japan), the largest optical telescope in Japan, and successfully detected the Crab pulsar \citep{Hashiyama+24}.
In addition, similar to \citet{Kanbach+08}, we detected optical SPs from the Crab pulsar during the MP phase. SPs in the IP component were also detected intermittently, although with a lower intensity.

This paper presents a study of the optical pulsed emission from the Crab pulsar, focusing on (i) the properties of optical SPs associated with GRPs, (ii) distributions of intensities and arrival timings of optical SPs, and (iii) the day-scale, short-term variation of the optical pulsed emission over a one-week observation period.
The article is organized as follows.
In Section \ref{sec:instrument}, we explain our high-time-resolution photometer IMONY upgraded from \citet{Nakamori+25}.
Section \ref{sec:observation} provides an overview of the observations conducted in both radio and optical bands.
Section \ref{sec:analysis} describes the analysis methods applied for radio and optical datasets, respectively.
The observational results are presented in Section \ref{sec:results}.
In Section \ref{sec:discussion}, we discuss the implications of the results.
Finally, Section \ref{sec:conclusion} summarizes our conclusions.

\section{Optical instrument IMONY} \label{sec:instrument}

\begin{figure}
	\begin{center}
		\includegraphics[width=8.5cm]{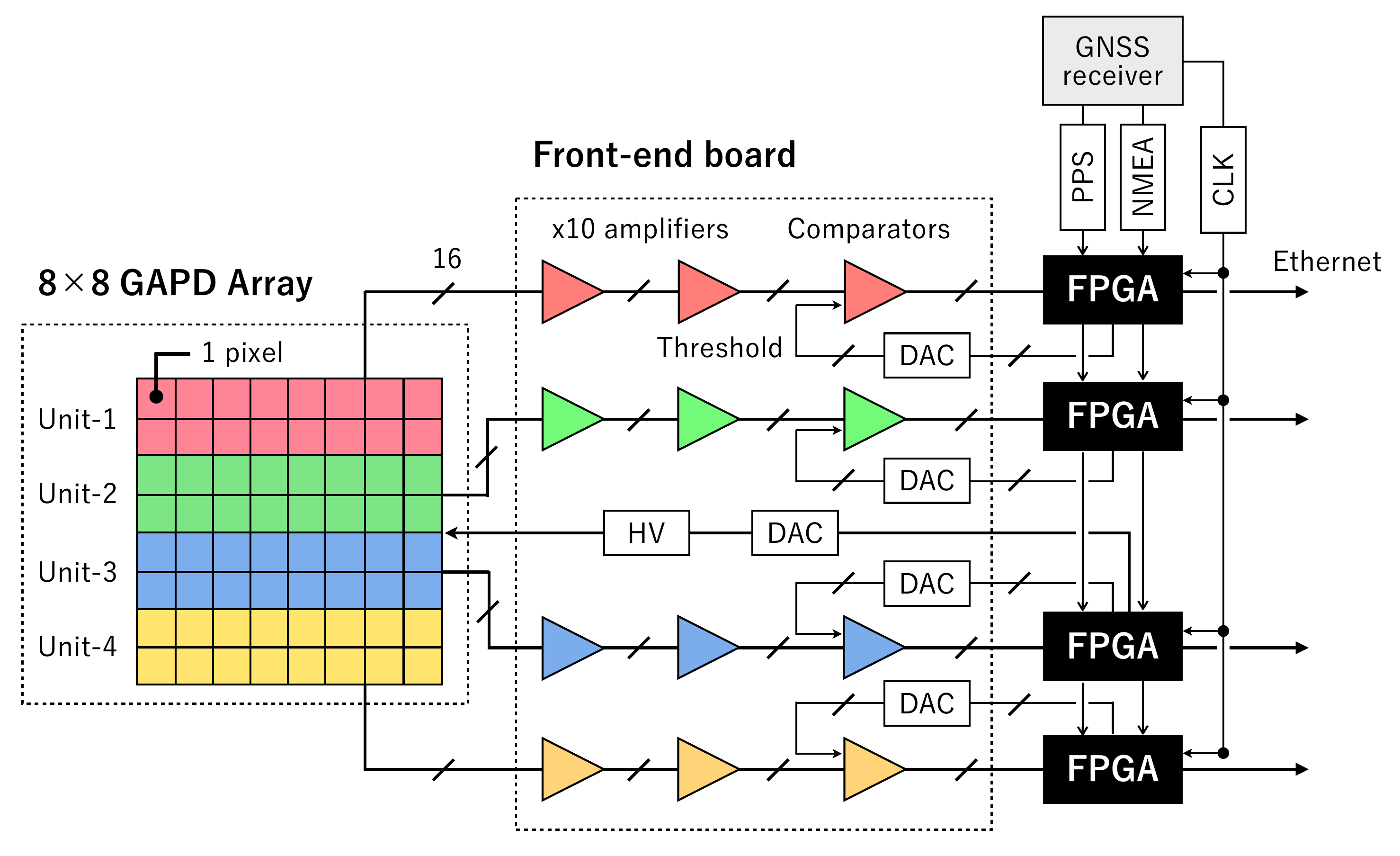}
	\end{center}
	\caption{A schematic view of the IMONY readout system.
    
    {Alt text: Block diagram of IMONY showing the signal flow from the $8\times8$ GAPD array to four FPGA boards via the front-end board.}

    }
	\label{fig:system_overview}
\end{figure}

Figure~\ref{fig:system_overview} shows the overall architecture of the IMONY readout system.  
The $8\times8$ GAPD array is divided into four units of 16 pixels each, with each unit processed by a dedicated acquisition module. 
Each module comprises a front-end board (FEB) and a Field-Programmable Gate Array (FPGA) board, as detailed in \citet{Nakamori+25}.
This work extends that single-module configuration to a four-module system, 
enabling full readout of the 64-pixel GAPD array.

Although the four modules operate independently, they are synchronized to ensure simultaneous data acquisition.  
A 10~MHz reference clock from the GNSS receiver is first fed into a clock buffer and then distributed equally to all four FPGAs to maintain phase alignment.  
In contrast, the pulse-per-second (PPS) signal and the National Marine Electronics Association-formatted sentences are provided to the first FPGA only and then relayed to the others in a daisy-chain manner via FPGA-to-FPGA connections (see Figure~\ref{fig:system_overview}).  
Each FPGA runs its own internal counters but shares the common reference timing through this mechanism, ensuring synchronization accuracy governed by the shared clock distribution.

Each module outputs time-stamped photon hit data via Ethernet using the SiTCP protocol \citep{Uchida+08}.  
When a photon is detected, the corresponding FPGA records the pixel address and the internal time-stamp counter value as a 64-bit word.  
This word typically includes a pixel ID (6~bits), a module ID (2~bits), and a 56-bit timestamp composed of the PPS counter and the 10~MHz clock counter.  
Photon hit data are transmitted only upon detection events; no data are recorded for inactive channels, thus eliminating the need for zero-suppression.
Consequently, the data volume scales linearly with the photon count rate, and is significantly smaller than conventional frame-based cameras.

On the backend computer, the data from each of the four modules are received and stored as separate binary files, resulting in four time-ordered data streams per observation.  
Since all modules share the same reference clock and are aligned in phase, the time-stamps across files can be merged offline to reconstruct the full $8\times8$ photon detection history with sub-microsecond accuracy.

\section{Observations} \label{sec:observation}

\begin{table*}
	\caption{Radio observation summary}
	\begin{tabular}{ccccccccc}
		\hline \hline
		Epoch				&    MJD     & 	Observatory 	& 	Start time	&	Duration	& 	Band		&	Centered frequency  		&	Bandwidth			&	Resolution  	\\
							&            &				    &	(UTC)	    &	(hour)	&	name	&	$\nu_\mathrm{c}$ (MHz)	&	$\Delta\nu$ (MHz)	&	(bit/sample)	\\ \hline
		February 7, 2024	&	60347    &  Usuda		    &	06:30:00	&	11.5		&	S		&	2258.0 				&	128				&	4			\\
							&	    	 &	                &	06:30:00    &	11.5		&	X		&	8438.0				&	128				&	4			\\
		February 8, 2024 	&	60348    &  Usuda		    &	06:30:00	&	11.5		&	S		&	2258.0 				&	128				&	4			\\
							&	         &	                &	06:30:00    &	11.5        &	X		&	8438.0				&	128				&	4			\\
		\hline
	\label{tab:obs_summary_radio}
	\end{tabular}
\end{table*}

\subsection{Radio}

We conducted radio observations using the Usuda telescope (Nagano, Japan) on February 7 and 8, 2024.
A summary of the radio observations is provided in Table~\ref{tab:obs_summary_radio}.
The telescope has a 64 m single-dish parabolic antenna and is located at 
(\timeform{138D21'54''}E, \timeform{36D07'44''}N).
Observations were performed simultaneously in two frequency bands: 2194–2322~MHz (S-band) and 8374–8502~MHz (X-band),  
each spanning 128~MHz and divided into four channels of 32~MHz bandwidth, sampled at 64~MHz.  
Data were digitized at 4 bits per sample in each channel.

These radio observations served two main purposes.  
The first was to determine the times of arrival (TOAs) of GRPs in barycentric dynamical time (TDB) at the Solar system barycenter.
Second, we used radio data to define the rotational phase $\phi$ of the Crab pulsar.
To compute $\phi$, we adopted the standard pulsar timing model with $\nu_0$, $\dot{\nu}_0$, and $\ddot{\nu}_0$, corresponding to the rotation frequency and its first and second derivatives at the reference time $t_0$:
\begin{equation}
    \psi\left(t\right) \approx \psi_0 + \nu_0 \left(t-t_0\right) + \frac{1}{2!} \dot{\nu}_0 \left(t-t_0\right)^2 + \frac{1}{3!} \ddot{\nu}_0 \left(t-t_0\right)^3,
    \label{eq:psi}
\end{equation}
where $\psi\left(t\right)$ is the number of pulsar rotations at $t=\mathrm{TOA}$, and $\psi_0$ is the initial rotation number at $t=t_0$.
Parameters in Equation (\ref{eq:psi}) are provided about monthly by the Jodrell Bank Observatory (JBO) for the Crab pulsar \citep{Lyne+93},
\footnote{\texttt{https://www.jb.man.ac.uk/pulsar/crab/crab2.txt}}
of which the nearest to our observation epochs was on February 15, 2024.
The rotational phase $\phi$ is then obtained as follows,
\begin{equation}
    \phi\left(t\right) = \psi\left(t\right) - \lfloor \psi\left(t\right) \rfloor.
    \label{eq:phi}
\end{equation}
$\psi_0$ in Equation (\ref{eq:psi}) is fine-tuned \textit{a posteriori} so that the average phase of the observed radio MP peak is zero.
Using $\psi_0$, we estimated the TOAs of the first MP after 00:00:00 TDB on the radio observation days and applied them to the optical analysis.
We note the different conventions in the definition of $\phi$ between the radio and optical/X-ray communities; the former uses $\phi=0$ but the latter do $\phi=1$ as the phase of the radio MP.
For comparison between radio and optical pulses we shift $+1$ for the phase of radio pulses.

\subsection{Optical}

\begin{table*}
	\caption{Optical observation summary}
	\begin{tabular}{ccccccccc}
		\hline \hline
		Epoch			    &     MJD           & 	Observatory 	&	Weather	    & 	Start time (UTC)	&	Duration (hour)    \\ \hline
		February 5, 2024	&     60345  	    &   Seimei		    &	Fine	    &	11:39:09	&	1.26	    \\
		February 6, 2024	& 	  60346  	    &   Seimei		    &	Cloudy	    &	10:30:24	&	0.91	    \\
		February 7, 2024	&	  60347  	    &   Seimei		    &	Cloudy	    &	12:02:36	&	1.15	    \\
		February 8, 2024	& 	  60348  	    &   Seimei 		    &	Cloudy	    &	11:23:38	&	0.79	    \\
		February 9, 2024	&	  60349  	    &   Seimei		    &	Fine	    &	09:53:46	&	2.06	    \\
		February 10, 2024	&	  60350  	    &   Seimei		    &	Fine	    &	09:59:57	&	2.25	    \\	
		February 11, 2024	&	  60351  	    &   Seimei		    &	Cloudy	    &	09:50:37	&	1.65	    \\
		\hline
	\label{tab:obs_summary_optical}
	\end{tabular}
\end{table*}

We installed the high-speed photometer IMONY on the Nasmyth focus of the Seimei telescope and observed the Crab pulsar from 9:00 to 15:00 UTC on February 5--11, 2024.
The Seimei telescope is located at 
(\timeform{133D35'48''2}E, \timeform{34D34'36''8}N),
with an altitude of 354.8~m above a sea level.
The telescope has a primary mirror diameter of 3780~$\mathrm{mm}$, and its focal length at the Nasmyth focus is 22692~mm ($\mathrm{f/D}\sim6$).
With a monolithic $8\times8$ GAPD array with a 200~$\mathrm{\mu m}$ pitch, IMONY yields a FoV of approximately $14\arcsec.5$ ($1\arcsec.8$ per pixel).

A summary of the optical observation conditions is also provided in Table~\ref{tab:obs_summary_optical}.
The weather on February 5, 9 and 10 was mostly clear, while the other days experienced cloudy sky with occasional clear intervals.
Each observation cycle consists of a 10-minute exposure on the Crab pulsar, bracketed by short dark frames taken before and after to monitor the dark count rate and to calibrate instrumental noise.
Although IMONY is not actively temperature-controlled, the sensor temperature is continuously monitored.
The recorded temperatures showed no significant variations around a few degrees during the observations and the dark count rate has remained stable around 500~Hz per pixel.
At the end of each observation epoch, we also acquired flat frames using a $3\times3~\mathrm{mm^2}$ white LED lamp (ENB01-NHSD7-F1 (P3H8)) mounted inside the dome near the roof structure.
These flat frames are used to calibrate the non-uniformity sensitivity of the sensor to incident photons.

\section{Analysis} \label{sec:analysis}
\subsection{Radio}

\begin{table*}
	\caption{Timing parameters during our observation period}
	\begin{tabular}{lccc}
		\hline \hline
		Parameter											&	February 7, 2024									&	February 8, 2024						&	Source	\\ \hline
		Right ascension (J2000.0)									&	5$^\mathrm{h}$34$^\mathrm{m}$31$^\mathrm{s}$.97232	&	Fixed								&	JBO	\\
		Declination	(J2000.0)									&	+22$^\circ$00$^\prime$52$^{\prime\prime}$.0690			&	Fixed								&	JBO	\\
		Ephemeris										&	DE200											&	Fixed								&	JBO	\\
		TOA of the first MP (MJD)								&	60347.000000344641204								&	60348.000000204212963					&	This work	\\
		Frequency $\nu_0$ ($\mathrm{s^{-1}}$)					&	$29.5641750805\pm0.0000000005$						&	$29.5641433963\pm0.0000000004$			&	JBO	\\
		Frequency first derivative $\dot{\nu}_0$ ($\mathrm{s^{-2}}$)		&	$(-366727.76\pm0.49)\times10^{-15}$					&	Fixed								&	JBO	\\
		Frequency second derivative $\ddot{\nu}_0$ ($\mathrm{s^{-3}}$)	&	$-1.99\times10^{-20}$								&	Fixed								&	JBO	\\
		Dispersion Measure ($\mathrm{pc~cm^{-3}}$)				&	$56.7507\pm0.0002$								&	Fixed								&	This work	\\
		\hline
	\label{tab:timing_params}
	\end{tabular}
\end{table*}

To correct for the dispersion caused by interstellar plasma, we applied coherent de-dispersion to the radio data using our own implementation based on the method of \citet{Hankins+71}.  
Although a reference Dispersion Measure (DM) was provided by the JBO for February 15, 2024 \citep{Lyne+93}, considering a time variation of the DM, we independently determined the DM using bright GRPs detected simultaneously in both the S- and X-bands on February 7 and 8, 2024.
The optimal DM was derived by iteratively adjusting the DM in steps of $0.0001~\mathrm{pc~cm^{-3}}$ and minimizing the time lag between the S- and X-band GRP peaks using cross-correlation at 500 ns resolution.

Using the optimal DM, we performed coherent de-dispersion and obtained a de-dispersed light curve
with an integration time of $10~\mathrm{\mu s}$ for the S-band and $1~\mathrm{\mu s}$ for the X-band to match the expected pulse widths.  
We converted the TOAs from UTC to TDB using the \texttt{PINT} package \citep{Luo+21}.
We then computed the rotational phase $\phi$ of the Crab pulsar using Equations (\ref{eq:psi}) and (\ref{eq:phi}), and determined the TOAs of the first MP on each observation epoch by constructing the radio phaseogram.
The timing parameters used for our analysis are summarized in Table \ref{tab:timing_params}.

\subsection{Optical}

We applied a two-step quality cut to remove low-quality data caused by atmospheric conditions and electronic noise \citep{Nakamori+25}. 
First, we removed time intervals with significant fluctuations in the photon count rate, which are indicative of transient cloud passages or instrumental spike noise.
After the rate-based quality cut, we performed a secondary the segment-based quality cut.
We converted the TOAs of optical photons from UTC to TDB and calculated rotational phase $\phi$ using Equations (\ref{eq:psi}) and (\ref{eq:phi}).
The timing parameters were adopted as those on February 8, 2024 (Table \ref{tab:timing_params}), corresponding to the midpoint of the optical observing sessions.
We then generated an optical phaseogram every 60 seconds and excluded segments with a peak signal-to-noise ratio (S/N) below 20.
The background noise was defined as the off-pulse baseline count level, representing the average photon count within the off-pulse interval between 0.7729 and 0.8446 in the phaseogram \citep{Slowikowska+09}.
Even under uniformly thin-cloud conditions, the segment-averaged baseline count can appear relatively stable.
However, the pulse detectability can still degrade due to attenuation and PSF blurring by thin clouds, resulting in the peak S/N falling below our threshold of 20.
Therefore, low-S/N segments were rejected primarily because of reduced pulse detectability.
After applying quality cuts, we performed calibrations by subtracting dark frames and applying flat-field corrections.  
Although dark frames were taken immediately before and after each observation run, we used the former to remove the instrumental noise.
In addition, flat-field correction was applied to compensate for pixel-to-pixel sensitivity variations.
Flat frames were background-subtracted using dedicated flat-dark frames, and the relative detection efficiency of each pixel was calculated from the ratio of its photon count to the array-wide average.
Photon counts were then normalized accordingly to correct for non-uniform sensor response.

\section{Results} \label{sec:results}

\subsection{Radio}

We performed test coherent de-dispersion on one hour observation data taken from 08:00:00 to 09:00:00 UTC on 8 February, 2024, identified two GRPs detected simultaneously in both the S- and X-bands.
The observed DMs for the first and second GRPs were determined to be $56.7505\pm0.0002~\mathrm{pc\;cm^{-3}}$ and $56.7508\pm0.0002~\mathrm{pc\;cm^{-3}}$, respectively.
Averaging these values, we adopted an observed DM of $56.7507\pm0.0002~\mathrm{pc\;cm^{-3}}$.
The corresponding timing error due to this DM uncertainty is $0.16~\mathrm{\mu s}$, which is significantly smaller than the integration time used in the radio light curves.

We determined the TOA of the first MP from the S-band phaseogram.
As the Crab pulsar was not detected in the X-band of the Usuda telescope, consistent with previous studies \citep{Nakamori+25}, we used the S-band data alone.
The TOAs of the first MP on each epoch are listed in Table \ref{tab:timing_params}.



Furthermore, we identified TOAs of GRPs in both the S- and X-bands.
A pulse width of a GRP in the S- and X-bands is an order of $\mathrm{\mu s}$, so the received anntena's power mostly comes from the radio emissions of the Crab Nebula and its pulsar wind nebula.
Thus, we computed S/N for a GRP by regarding the average power as the background and the standard deviation of the power as the $1\sigma$ level noise.
In this study, referring to \citet{Mikami+16}, who used the same telescope, we set a detection threshold of $7\sigma$ for the S-band and $9\sigma$ for the X-band, respectively.

\subsection{Optical}

\begin{figure}
	\includegraphics[width=8.5cm]{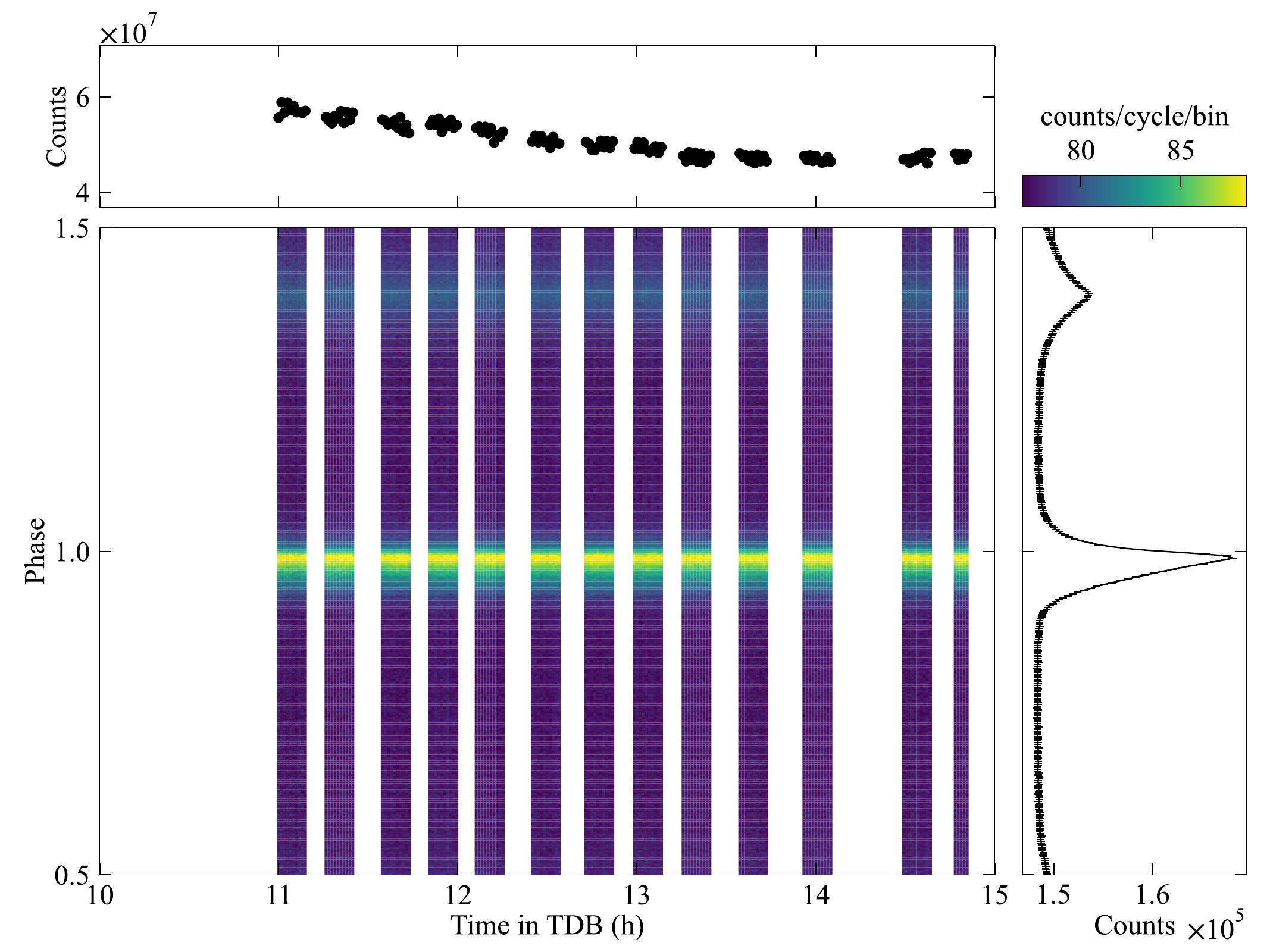}
	\caption{Observation results after the two-step quality cut on February 9, 2024. Approximately 80\% of data was survived after the cut. The top-left panel shows the photon count rate over 64 pixels, and the bottom-left panel displays the optical 2D phaseogram. The time intervals in both plots are 60 seconds. The bottom-right panel illustrates the phaseogram with a bin width of $100~\mathrm{\mu s/bin}$, also summed over 64 pixels. 
    
    {Alt text: A two-dimensional histogram showing a relationship among the barycentric dynamical time (horizontal axis), rotational phase (vertical axis), and total photon counts accumulated over 60 second intervals. The projections onto the phase and time axes are displayed above and to the right of the histogram, respectively.}
    
    }
	\label{fig:20240209_2DLC}
\end{figure}

February 9, 2024 was selected as a representative night for optical analysis due to its favorable observing conditions, including clear skies and stable seeing throughout the night.
These conditions enabled optical photometry with high S/N.
Figure~\ref{fig:20240209_2DLC} summarizes the optical observation results on that date. 
The bottom-left panel presents a two-dimensional phaseogram (2D phaseogram), illustrating the relationship among the elapsed time in TDB (horizontal axis), rotational phase (vertical axis), and photon count rate (color scale).
The top-left panel displays the photon counts integrated over 64 pixes in 60 second intervals, revealing a gradual decline over time. This smooth trend is attributed to the increasing airmass (zenith angle) as the target's altitude decreased during the observation.
The bottom-right panel shows the phaseogram integrated over time, showing a clear pulsed profile aligned with the Crab pulsar's rotational phase.
The approximately 5-minute gap visible between two observation runs corresponds to the acquisition time for dark frames.
\footnote{During these intervals, precautionary imaging with the Seimei telescope’s main camera TriCCS was also performed. The data were not used in this paper.}

\begin{figure}
	\includegraphics[width=8.5cm]{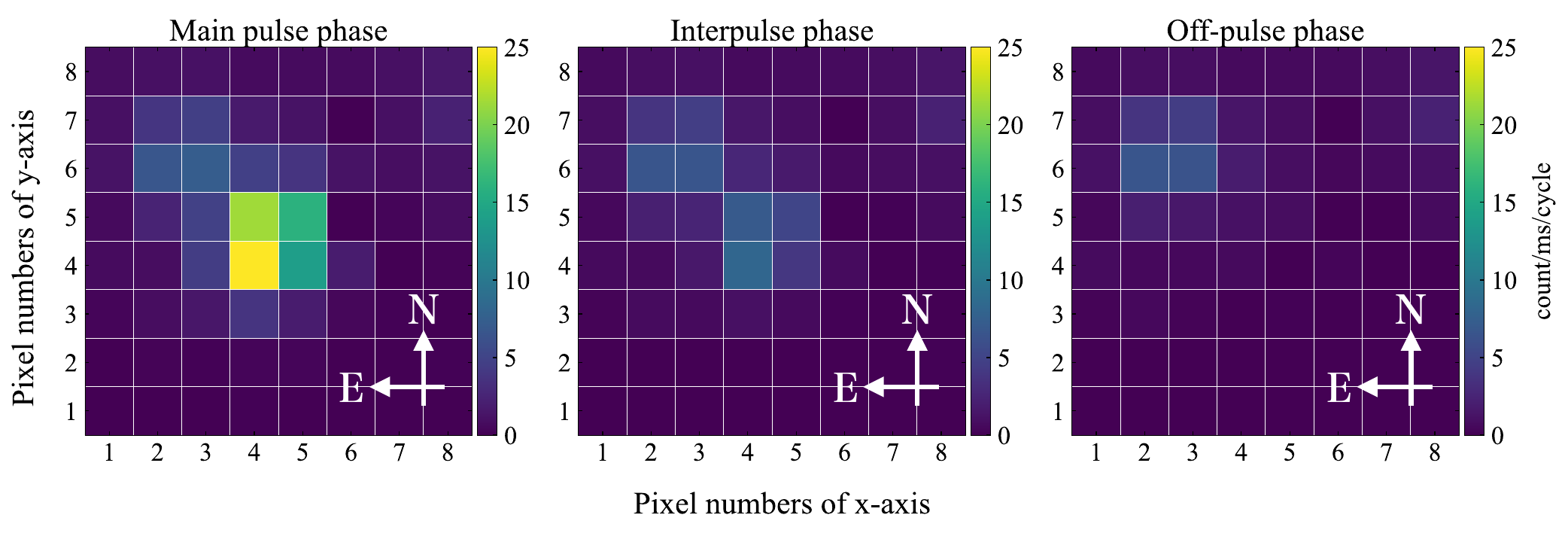}
    \caption{
    Snapshot images corresponding to three rotational phase intervals of the Crab pulsar: MP (left), IP (middle), and off-pulse (right). Each frame is integrated over 10 minutes. The bright spot during the MP phase represents the pulsar. A nearby reference star (2MASS J05343217+2200560) is also visible to the northeast. North is up and East is to the left.
    
    {Alt text: Snapshot images of the Crab pulsar region in the MP, IP, and the off-pulse phases.}
    
    }
    \label{fig:20240209_snapshots}
\end{figure}

We present three types of snapshot images in Figure~\ref{fig:20240209_snapshots}, corresponding to the rotational phase of the MP (left), IP (middle) and off-pulse (right),
each integrated over 10 minutes.
We adopted the phase intervals for MP and IP as 0.8500 to 1.1500 and 1.3500 to 1.5500, respectively.
Also, the off-pulse phase was defined as 0.7729--0.8447 \citep{Slowikowska+09}.
A brightness variation is clearly visible at the center of the FoV, corresponding to the location of the Crab pulsar.
The PSF of the pulsar appeared as a $2\times2$ pixel spot under the seeing conditions of the night.
Therefore, we summed photon counts over $4\times4$ pixels (equivalent to $7\arcsec.2\times7\arcsec.2$) to cover the entire PSF in the subsequent analysis.
Owing to the expanded FoV of IMONY compared to its prototype \citep{Nakamori+25}, we were also able to image a reference star, 2MASS J05343217+2200560 (magnitude 16.64), located to the northeast of the Crab pulsar.

\begin{figure*}
	\begin{center}
		\includegraphics[width=15cm]{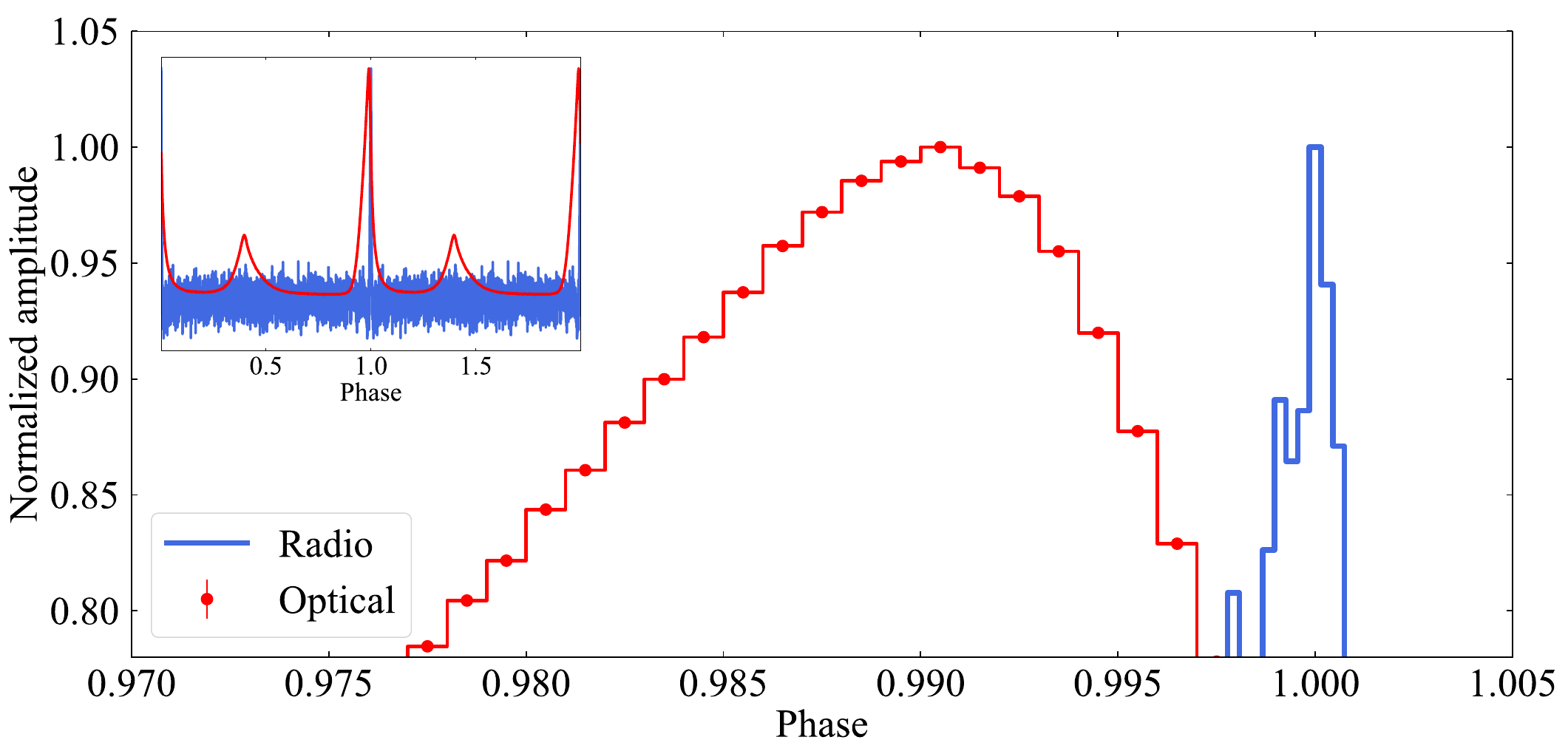}
	\end{center}
	\caption{Comparison between the optical and radio pulse profiles. The red line shows the optical phaseogram divided into 1000 phase bins per rotation. The blue line shows the radio phaseogram recorded on February 7 and 8, 2024, in the S-band of the Usuda telescope. Phase 1.0 is aligned with the radio peak. The inset shows the full optical and radio profiles across two rotations.

    {Alt text: Close-up optical and radio phaseograms of the Crab pulsar around the MP peak, from phase 0.970 to 1.005, with the full-range profiles shown in the inset panel.}
    
    }
	\label{fig:total_phaseogram}
\end{figure*}

\begin{figure}
	\includegraphics[width=8cm]{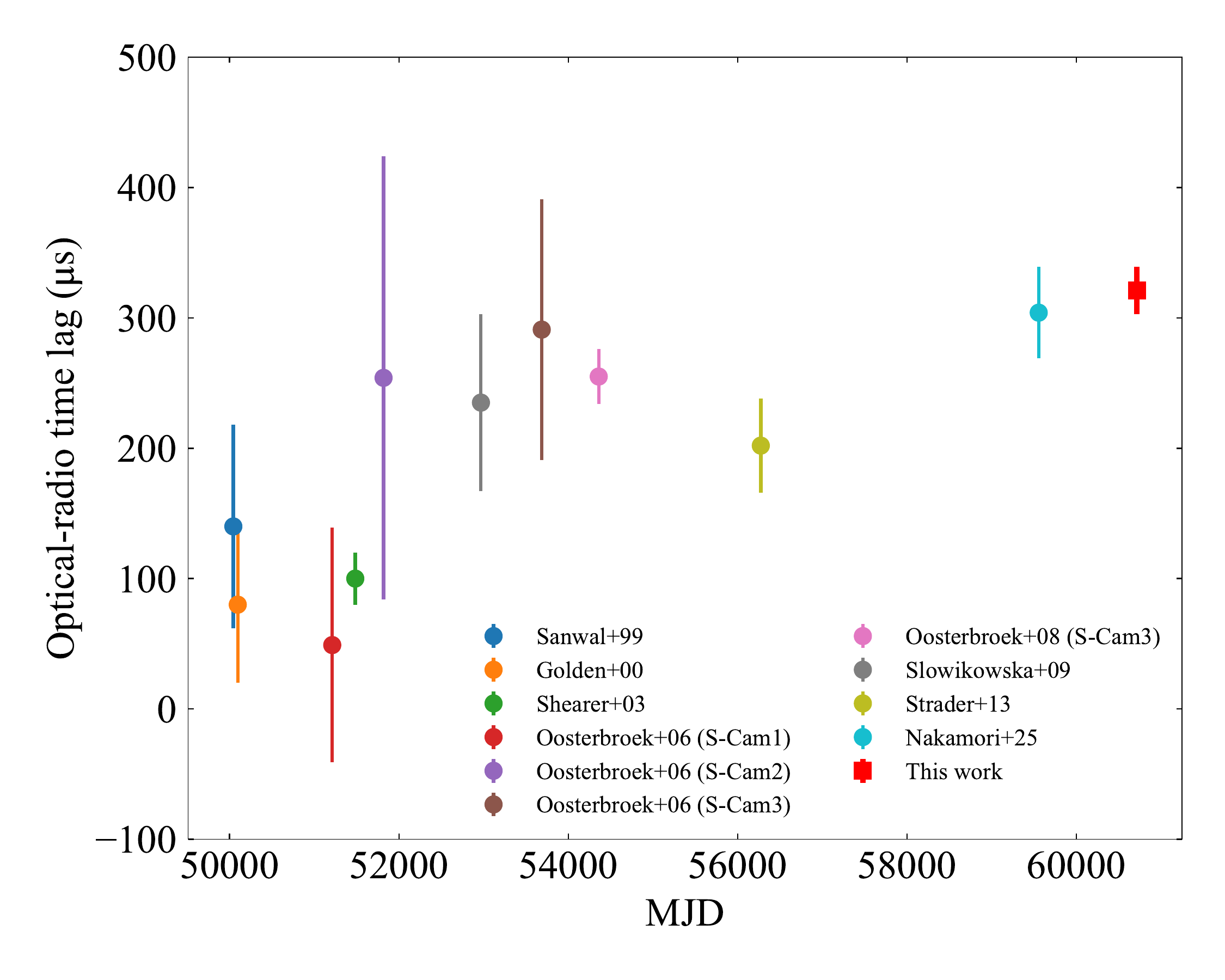}
	\caption{Comparison of the optical-radio time lags between our result and previous observations. Our result is consistent with \citet{Oosterbroek+06}, \citet{Slowikowska+09}, and \citet{Nakamori+25} within the $1\sigma$ error range.

    {Alt text: Scatter plots with error bars, showing the comparison of optical-radio lags obtained in our observations and in previous studies. The horizontal axis spans 1999--2024 in MJD, and the vertical axis ranges from $-100~\mathrm{\mu s}$ to $500~\mathrm{\mu s}$.}
    
    }
	\label{fig:CompLag}
\end{figure}

Folding the calibrated light curves extracted from the $4\times4$ region, we constructed an optical phaseogram of the Crab pulsar over the period from February 5 to 11, 2024, as shown in Figure~\ref{fig:total_phaseogram}.
The phaseogram was divided into 1000 phase bins per rotation, corresponding to the temporal resolution of approximately $34~\mathrm{\mu s/bin}$.
The time lag between both optical and radio peaks (optical-radio time lag) was measured as $\Delta\phi = 0.0095 \pm 0.0005$, corresponding to $\Delta t=321\pm18~\mathrm{\mu s}$.
The total uncertainty of $18~\mathrm{\mu s}$ was calculated by adding in quadrature three independent components: $0.16~\mathrm{\mu s}$ from the DM, $5~\mathrm{\mu s}$ from half of the radio integration time, and $17~\mathrm{\mu s}$ from half of the optical phaseogram bin width.
Figure \ref{fig:CompLag} compares our result with the values reported by various authors.
Our value is consistent with previous observations \citep{Oosterbroek+06, Slowikowska+09, Nakamori+25}, while it does not agree with other measurements \citep{Sanwal+99, Golden+00, Shearer+03, Oosterbroek+08, Strader+13}.

\begin{figure}
	\includegraphics[width=8.5cm]{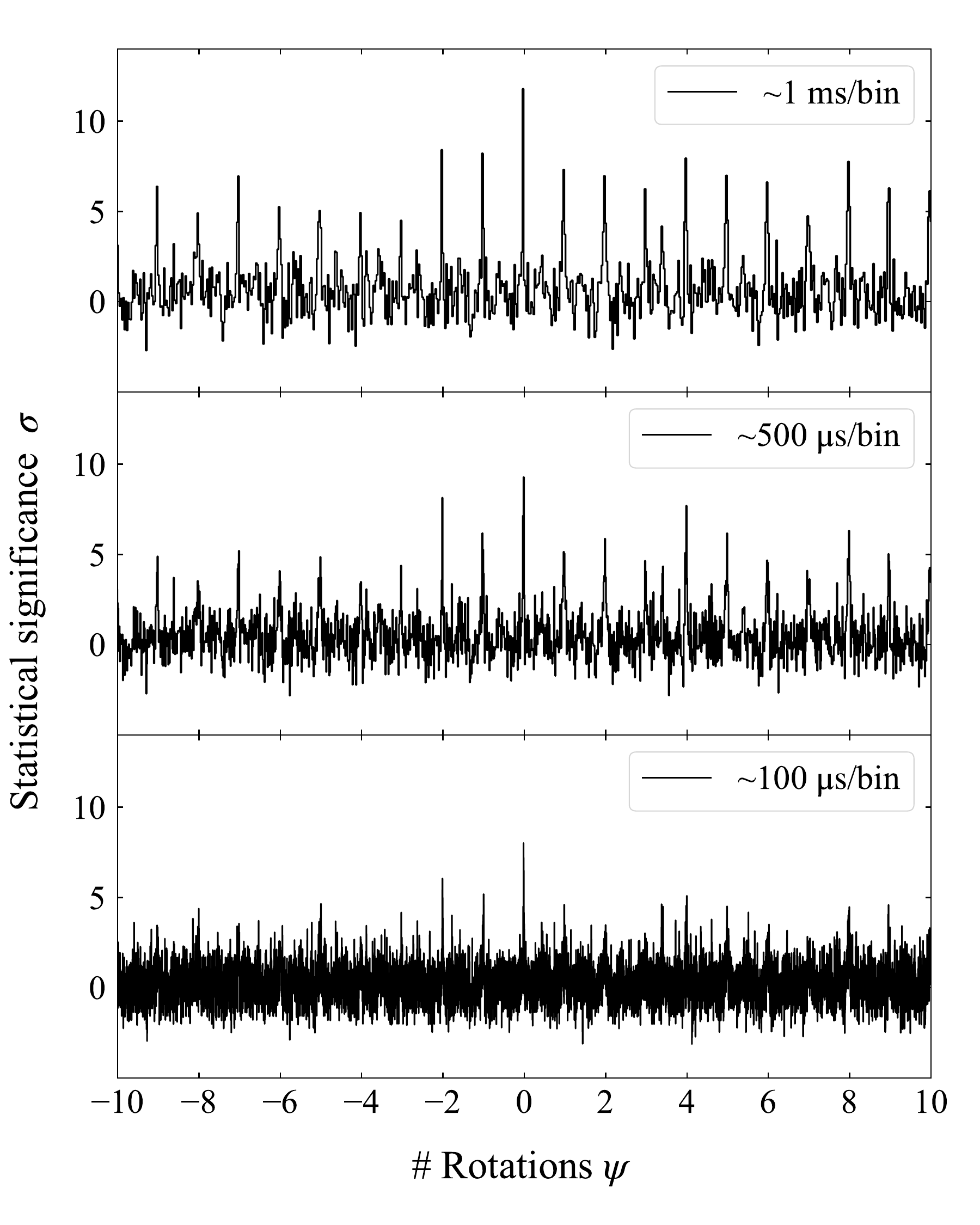}
	\caption{Observed optical SPs on February 9, 2024, using IMONY mounted on the Seimei telescope. The top, middle, and bottom panels show light curves with different time bin widths. The horizontal axis represents the sequence of pulsar rotations, while the vertical axis indicates the detection significance. Light curves are shown over 20 rotation cycles, corresponding to approximately $680~\mathrm{ms}$. The statistical significance of MPSPs was about $5\sigma$ on average for time bins of at least approximately $500~\mathrm{\mu s/bin}$.

    {Alt text: Three line graphs of the Crab pulsar optical light curve over 20 rotations, with $1~\mathrm{ms/bin}$ (top), $500~\mathrm{\mu s/bin}$ (middle), and $100~\mathrm{\mu s/bin}$ (bottom).}
    
    }
	\label{fig:optical_sps}
\end{figure}

Thanks to the high-time resolution of IMONY and the large aperture of the Seimei telescope, we successfully detected optical SPs from the Crab pulsar.
Figure \ref{fig:optical_sps} shows time series of detection significance over 20 consecutive pulsar rotations, evaluated using different time bin widths.
The vertical axis represents the significance of each pulse, calculated relative to the Poisson fluctuations in the off-pulse interval defined between 0.7729 and 0.8446 \citep{Slowikowska+09}.
Peaks with high significance appear at every phase of the main pulse, particularly in the top and middle panels, confirming the successful detection of optical main pulse SPs (MPSPs). These detections remain robust down to at least approximately $500~\mathrm{\mu s/bin}$, where the average significance reaches approximately $5\sigma$. In contrast, individual interpulse SPs (IPSPs) are less prominent, with significance typically around $3\sigma$.

To characterize these SPs in more detail, we focused on four waveform parameters for each optical MPSP. These parameters were the integrated photon count (optical fluence), the peak count, the peak timing, and the pulse width defined as the Full Width at Half Maximum (FWHM). 
In this analysis, we used light curves binned at approximately $500~\mathrm{\mu s/bin}$, which provided sufficient S/N and photon counting statistics.
We fitted each MPSP waveform with a Gaussian function to estimate these parameters.
We note that the average profile of the MP is not Gaussian in shape. However, for individual MPSPs with limited photon counts, Gaussian model provided a stable and consistent approximation for pulse fitting.
We evaluated the goodness of fit using the $\chi^2$ test, adopting a 5\% confidence level, and excluded MPSPs with $p$-value less than 0.05.
We also removed events that showed poor convergence, defined by a relative amplitude exceeding than 10.

The detection efficiency of optical MPSPs varied from night to night, which is consistent with the weather conditions described earlier. On February 5, 2024, when the weather was favorable, the efficiency reached 77\%, whereas it dropped to 39\% on February 8, when the conditions were poorer. 
The detection efficiency was defined as the fraction of MPs that were successfully identified as MPSPs, among those occurring during time intervals that survived both stages of the quality cuts.

\begin{figure}
	\includegraphics[width=8cm]{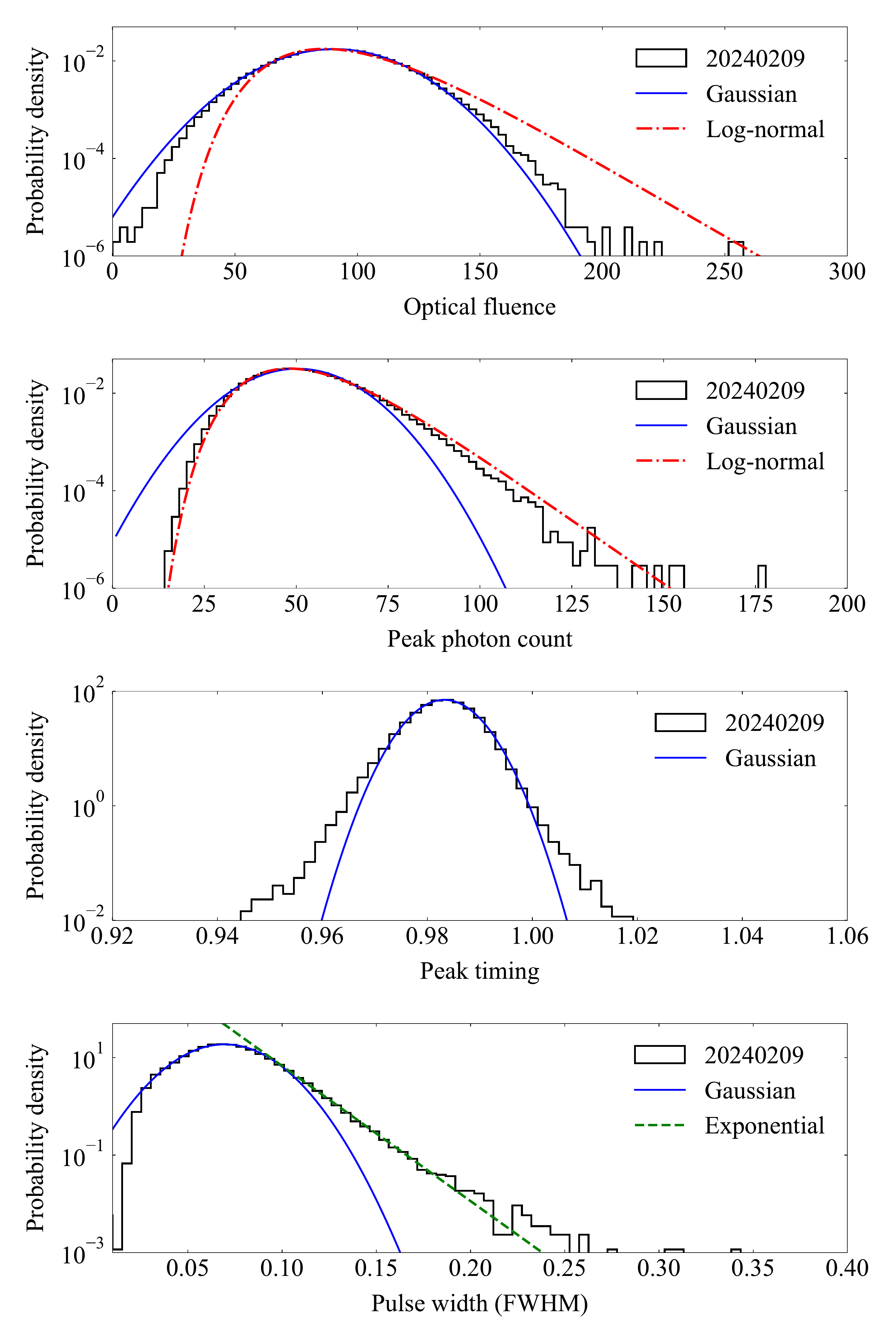}
	\caption{Distributions of four waveform parameters for optical MPSPs observed on February 9, 2024. Fluence (top panel), peak photon count (second panel), peak timing (third panel), and pulse width in terms of FWHM (bottom panel). The blue (solid), red (dash--dot), and green (dashed) lines represent the fitting results using the Gaussian, log-normal, and exponential functions, respectively.

    {Alt text: Four histograms showing the distributions of the four pulse parameters of optical MPSPs, with their fitting results also displayed.}
    
    }    
	\label{fig:20240209_ParamsSP}
\end{figure}

The distributions of these parameters on February 9, 2024, are shown in Figure \ref{fig:20240209_ParamsSP}.
The top panel shows the distribution of the optical fluence, which is better described by a Gaussian distribution rather than a log-normal distribution.
The second panel presents the peak photon count distribution.
In contrast to the fluence, the peak count more closely follows a log-normal distribution.
This difference suggests that brighter pulses are narrower and exhibit higher peaks. 
This behavior resembles that of GRPs from the Crab pulsar reported by \citet{Bhat+08}.
The third panel displays the distribution of the peak timing,
where the horizontal axis is given in the rotational phase.
This distribution is approximately Gaussian around the peak, but exhibits the tail component on both sides.
Finally, the bottom panel shows the distribution of the pulse width,
expressed as the FWHM and also in phase units.
The peak region is well fitted by a Gaussian, while the right-hand tail follows an exponential tail.

\section{Discussion} \label{sec:discussion}

This paper presents the first statistical characterization of optical single-pulse parameters from the Crab pulsar, based on high-time-resolution observations using the 64-channel version of IMONY.
While the correlation between optical pulses and GRPs has been previously studied by other instruments, this work represents the first such investigation conducted with IMONY.
The initial instrument performance and its SP detection capability were briefly reported in \citet{Hashiyama+24}, and this is the first scientific analysis using IMONY data to explore both the intrinsic properties of optical SPs and their connection to GRPs.

\subsection{Validation of the optical enhancement}
\begin{figure}
	\includegraphics[width=8cm]{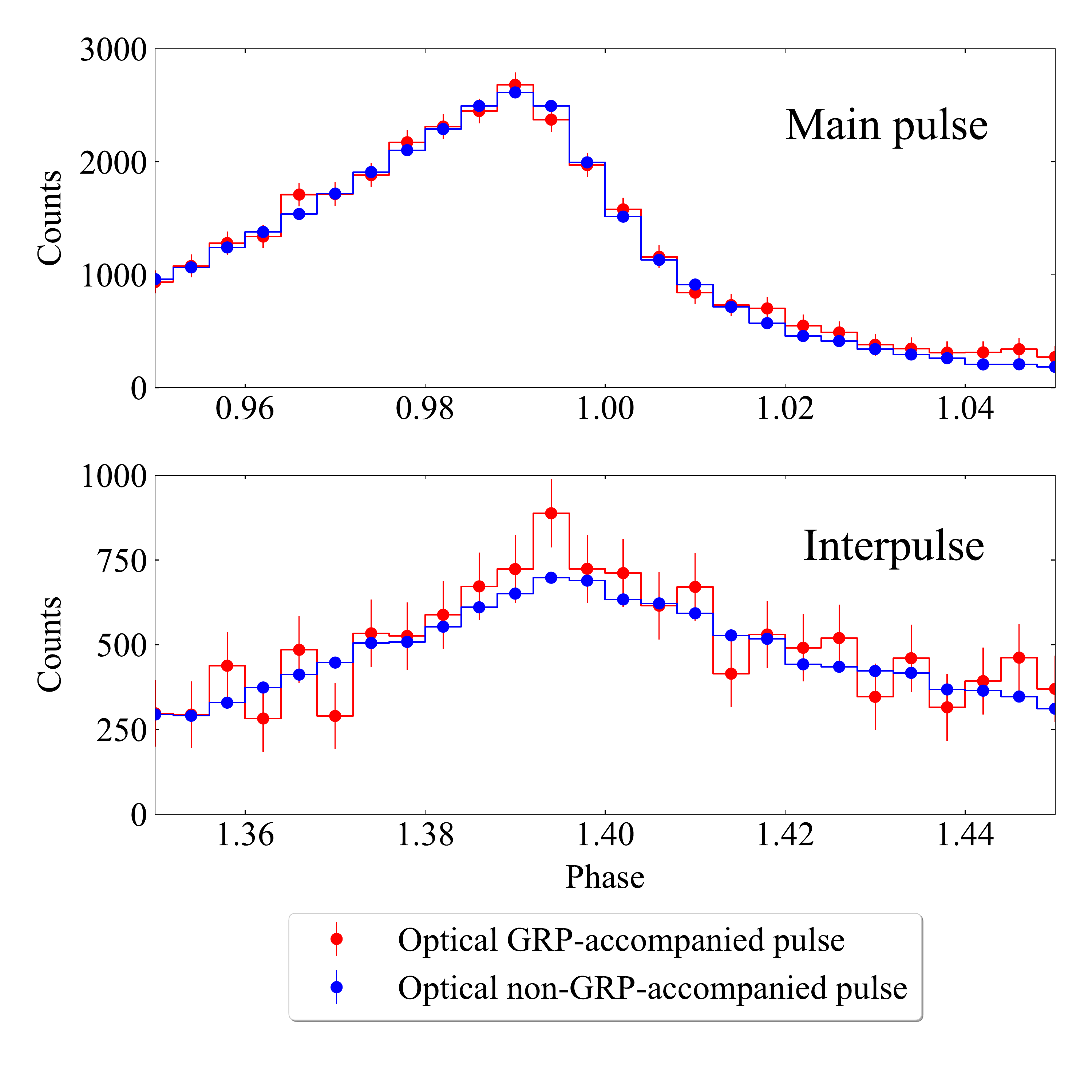}
	\caption{Stacked optical light curves in the MP (top panel) and IP (bottom panel) phases. No significant enhancement was observed, likely due to limited statistics.

    {Alt text: Close-up optical phaseograms around the MP (top) and IP (bottom) phases. Each panel contains two step plots with error bars, corresponding to the GRP-accompanied and non-GRP-accompanied optical pulse profiles.}
    
    }
	\label{fig:optical_enhancement}
\end{figure}

The optical MPs accompanied by GRPs are enhanced by approximately 3\% on average \citep{Shearer+03,Strader+13}.
Using our simultaneous optical and radio datasets, we examined this enhancement effect.
During approximately 22-hour observing runs, we identified a total of 9569 MPGRPs and 564 IPGRPs in the S-band (above the $7\sigma$ threshold), and 120 MPGRPs and 284 IPGRPs in the X-band (above $9\sigma$).
We set both thresholds referring the criteria used by \citet{Mikami+16}, with the S-band threshold set more strictly than the reference to account for contamination by fake GRPs, excluding MPGRPs and IPGRPs.
Among these MPGRP and IPGRP detections, 638 rotation cycles were simultaneously detected with optical SPs.
Figure \ref{fig:optical_enhancement} shows the stacked light curves for these events.
In \citet{Strader+13}, the folded pulse profile for non-GRP-accompanied pulses was constructed from optical pulses within the 40 rotations before and after each GRP-accompanied pulse.
Following \citet{Strader+13}, we also adopted the same definition.
We then calculated the enhancement ratios $R_i$ ($i=\mathrm{MP,\;IP}$) for each of the MP and IP peaks from the fraction of averaged peak intensities of GRP-accompanied $I_\mathrm{peak,\;GRPacc}$ and non-GRP-accompanied pulses $I_\mathrm{peak,\;nGRPacc}$.
\begin{equation}
    R_i = \frac{I_\mathrm{peak,\;GRPacc} - I_\mathrm{peak,\,nGRPacc}}{I_\mathrm{peak,\,nGRPacc}}\;\;\;(i=\mathrm{MP,\;IP})
\end{equation}
The enhancement ratio was $2.58\pm4.19\%$ at the MP peak and $27.3\pm14.1\%$ at the IP peak.
From our data, we estimated the $2\sigma$ upper limits of the optical enhancement to be $11.0\%$ and $55.5\%$ at both peaks.




Unfortunately, we could not confirm significant optical enhancement for either of the components.
This is likely due to insufficient statistics, with our sample size much smaller than in previous studies \citep{Shearer+03,Strader+13}.
In our observations, the effective duration of simultaneous optical and radio observations was limited to about one hour due to cloud cover, resulting in fewer than 1000 optical GRP-accompanied pulses.
By contrast, \citet{Strader+13} collected 7205 simultaneous events and detected optical enhancement with grater than $7\sigma$ significance.
Given the GRP detection rate was at least 15 events per minutes in the S-band from our radio observation, we therefore an additional 7 hours of effective simultaneous observation will be necessary to detect the optical enhancement with IMONY.
If the enhancement effect is detected in future observations with IMONY, detailed discussions of the variability of the pair production rate in the pulsar magnetosphere (e.g., \citet{Shearer+03}) and of the number of merged plasmoids (e.g., \citet{Enoto+21}) will be possible.







\subsection{Optical peak timing stability over one week}
Long-term modulations in the optical phaseogram of the Crab pulsar have been suggested on a yearly timescale \citep{Karpov+07}.
In this study, we investigated the day-scale, short-term stability of the optical pulsed emission over one week, using one of our longest nightly datasets.
We adopted the phaseogram on February 8, 2024, midpoint of our observation period, as the reference profile.
Each phaseogram was fitted with a multi-Lorentzian model developed by \citet{Zampieri+14}, and the peak timing of the MP was extracted.
The error of the peak timing includes several contributions, such as fluctuation in the radio peak timing $\sigma_\mathrm{r}$, variations caused by the DM modulation $\sigma_\mathrm{DM}$, the extrapolation error in the pulsar rotation frequency $\sigma_\mathrm{\nu}$, and the fitting uncertainty of the peak itself $\sigma_\mathrm{f}$.
The radio peak error $\sigma_\mathrm{r}$ was conservatively taken as half of the radio integration time, $10~\mathrm{\mu s}$.
The DM-induced timing error $\sigma_\mathrm{DM}$ due to the uncertainty in the DM was estimated to be $0.16~\mathrm{\mu s}$.
The error $\sigma_\mathrm{\nu}$ from the extrapolating the rotation frequency grows with the time offset from February 15, 2024, reaching a maximum of $10.7~\mathrm{\mu s}$ on February 5 and a minimum of $3.6~\mathrm{\mu s}$ on February 11.
The total timing uncertainty between two observing days (day-1 and day-2) was then calculated as
\begin{equation}
	\Delta\tau = \sqrt{\left(\sigma_\mathrm{f_1}^2+\sigma_\mathrm{f_2}^2\right) + \left(\sigma_{\nu_1}^2+\sigma_{\nu_2}^2\right)},
\end{equation}
where $\sigma_\mathrm{f_1}$ and $\sigma_\mathrm{f_2}$ represent the peak fitting errors for day-1 and day-2, respectively.
The common systematic errors $\sigma_\mathrm{r}$ and $\sigma_\mathrm{DM}$ were not included in this relative error, as they cancel out in pairwise comparisons.

\begin{figure}
	\begin{center}
		\includegraphics[width=8cm]{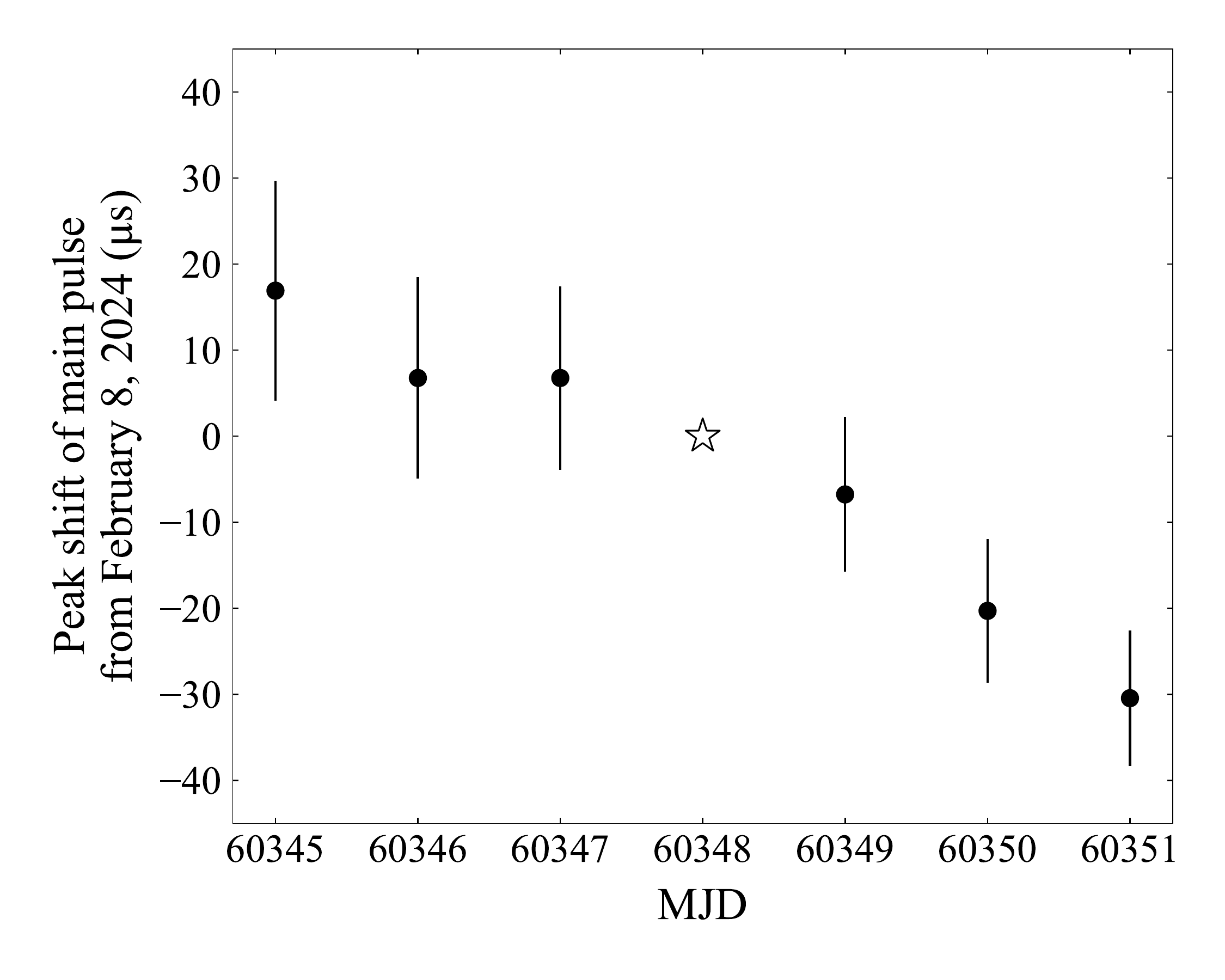}
	\end{center}
	\caption{Time variation of the peak timing of the MP relative to the reference profile. The star marks the reference point on 8 February, 2024. The peak lag increases over time, reaching a maximum of $30~\mathrm{\mu s}$ on MJD 60351 (February 11, 2024), with a statistical significance of $3.9\sigma$.

    {Alt text: Scatter plots with error bars showing the time variations of optical MP peak timing differences relative to the reference timing on 8 February, 2024. The timing differences move from $+20~\mathrm{\mu s}$ to $-30~\mathrm{\mu s}$ between 5 and 11 February, 2024.}
    
    }
    \label{fig:timing_diff}
\end{figure}

We show the time variation of the peak timing in Figure \ref{fig:timing_diff}, and the numerical results are also summarized in Table \ref{tab:timing_diff}.
The timing differences $\tau$ tend to be large over time, reaching a maximum deviation of $30~\mathrm{\mu s}$ from February 8 to 11, with a statistical significance of $3.9\sigma$.
The observed timing drift corresponds to a light travel distance of $c\tau\sim9.1~\mathrm{km}$, which is approximately 0.006 times the light cylinder radius of the Crab pulsar of about 1600~km.

One possible explanation for the observed timing drift is a pulsar glitch.
To test this possibility, we traced the rotation period at each epoch using our optical data and the method of \citet{Zampieri+14}; however, we did not detect any rapid changes in the rotation period.

Another possible interpretation is a drift of the optical emission region.
Assuming that the optical emission originates from the Outer Gap (OG) (e.g., \citet{Cheng+86}), this displacement is consistent with the theoretically expected OG size of a few hundred km \citep{Takata+04}.
The observed spatial displacement also could be attributed to fluctuation of the Y-point around the light cylinder or caustic in the pulsar magnetosphere.
Furthermore, the spatial distance of 9.1~km is much larger than the typical current sheet thickness of 30~m \citep{Philippov+19}, implying the need for plasmoids that are sufficiently developed by merging.
These merged plasmoids may play a role in the bright emissions such as GRPs.


\begin{table}
	\caption{Timing differences of MPs in our observation period}
	\begin{tabular}{cccc}
		\hline \hline
			Epoch in MJD & Difference & Error & Significance \\
            & $\tau$ ($\mathrm{\mu s}$) & $\Delta\tau$ ($\mathrm{\mu s}$) & $\sigma$ \\
			\hline
			60345 & 17 & 13 & 1.3 \\
			60346 & 6.8 & 12 & 0.57 \\
			60347 & 6.8 & 11 & 0.63 \\
			60348 & -- & -- & -- \\
			60349 & $-$6.8 & 9.0 & 0.75 \\
			60350 & $-$20 & 8.4 & 2.4 \\
			60351 & $-$30 & 7.9 & 3.9 \\
			\hline
	\label{tab:timing_diff}
	\end{tabular}
\end{table}

\section{Conclusion} \label{sec:conclusion}

In this paper, we reported the results of nightly optical observations of the Crab pulsar using upgraded IMONY mounted on the 3.8~m Seimei telescope.
On two of those nights, we also performed simultaneous radio and optical observations.

The peak of the optical phaseogram led the radio peak by $321\pm18~\mathrm{\mu s}$, which was consistent with previous studies. 
Furthermore, we successfully detected optical SPs with a significance of $5\sigma$ for MPSPs and approximately $3\sigma$ for IPSPs with a time bin of $500~\mathrm{\mu s}$.
In this paper, we focused on the MPSPs and examined the distributions of four pulse parameters: the optical fluence, peak photon count, peak timing, and pulse width (FWHM).
The optical fluence followed a Gaussian distribution, peak photon count followed a log-normal distribution, peak timing followed a Gaussian distribution with a trailing tail, and the pulse width followed a Gaussian distribution with an exponential cut-off.


We attempted to verify the optical enhancement associated with GRPs; however, no statistically significant enhancement was found in our data due to the limited duration of simultaneous observations.
The measured enhancement ratio was $2.58\pm4.19\%$ for the MP and a $27.3\pm14.1\%$ for the IP.
The lack of significance is likely due to insufficient statistics, as the number of optical pulses coincident with GRPs in our data was less than one-tenth of that reported by \citet{Strader+13}.
Based on the S-band detection rate of 15 events per minute, an additional approximately 7 hours of effective simultaneous observation time is estimated to be required to achieve a statistically significant detection of the optical enhancement.



Notably, one week of optical observations revealed that the average peak timing drifted by $30~\mathrm{\mu s}$ over three days, with a significance level of $3.9\sigma$.
This shift exceeds the estimated systematic errors and corresponds to a light travel distance of about $9.1~\mathrm{km}$, which is 0.006 times smaller than the light cylinder radius of the Crab pulsar.
Although a pulsar glitch could explain such modulation, we detected no rapid change in the rotation period. 
Therefore, we proposed that the observed timing shift may reflect a physical drift of the optical emission region.
Assuming the OG model, the spatial displacement is consistent with the expected OG depth of several hundred kilometers \citep{Takata+04}.
It may also be explained by fluctuations of the Y-point around the light cylinder and by variations in the magnetospheric caustics.
The observed spatial distance is far larger than the typical current sheet thickness, suggesting that the plasmoid merger and subsequent growth could account for the large displacement.
These developed plasmoids may contribute to bright emissions such as GRPs.

\begin{ack}
We thank Shinpei Shibata for fruitful discussions from a theoretical perspective on our observational results.
This work was supported by JSPS KAKENHI Grant Numbers 23K25890, 23K22538, and 22K03681, and by JST SPRING Grant Number JPMJSP2108.
\end{ack}


\end{document}